\documentclass[twocolumn]{aastex631}
\usepackage{graphicx}
\usepackage{amsmath}
\usepackage{CJK}
\usepackage{soul}
\usepackage{makecell}

\newcommand{\coo}{$\mathrm{C^{18}O}$}

\shorttitle{Disk Masses by \coo}
\shortauthors{D. Deng et al.}

\graphicspath{{./}{figures/}{figures/figures_archive/}}

\begin{document}

\title{\texttt{DiskMINT}: A Tool to Estimate Disk Masses with CO Isotopologues}

\author[0000-0003-0777-7392]{Dingshan Deng} 
\affiliation{Lunar and Planetary Laboratory, the University of Arizona, Tucson, AZ 85721, USA}

\correspondingauthor{Dingshan Deng}
\email{dingshandeng@arizona.edu}

\author[0000-0003-0522-5789]{Maxime Ruaud}
\affiliation{NASA Ames Research Center, Moffett Field, CA 94035, USA}
\affiliation{Carl Sagan Center, SETI Institute, Mountain View, CA 94043, USA}

\author[0000-0002-3311-5918]{Uma Gorti}
\affiliation{NASA Ames Research Center, Moffett Field, CA 94035, USA}
\affiliation{Carl Sagan Center, SETI Institute, Mountain View, CA 94043, USA}

\author[0000-0001-7962-1683]{Ilaria Pascucci}
\affiliation{Lunar and Planetary Laboratory, the University of Arizona, Tucson, AZ 85721, USA}


\begin{abstract}

CO is one of the most abundant molecules in protoplanetary disks, and optically thin emission from its isotopologues has been detected in many of them. 
However, several past works have argued that reproducing the relatively low emission of CO isotopologues requires a very low disk mass or significant CO depletion. 
Here, we present a \texttt{Python} code, \texttt{DiskMINT}, which includes gas density and temperature structures that are both consistent with the thermal pressure gradient, isotope-selective chemistry, and conversion of CO into $\mathrm{CO_2}$ ice on grain-surfaces.
The code generates a self-consistent disk structure, where the gas disk distribution is obtained from a Spectral Energy Distribution (SED)-derived dust disk structure with multiple grain sizes.
We use \texttt{DiskMINT} to study the disk of RU~Lup, a high-accreting star whose disk was previously inferred to have a gas mass of only $\sim 1.5\times10^{-3}\,M_\odot$ and gas-to-dust mass ratio of $\sim 4$.
Our best-fit model to the long-wavelength continuum emission can explain the total $\mathrm{C^{18}O}$ luminosity as well as the $\mathrm{C^{18}O}$ velocity and radial intensity profiles, and obtains a gas mass of $\sim 1.2\times10^{-2}\,M_\odot$, an order of magnitude higher than previous results.
A disk model with parametric Gaussian vertical distribution that better matches the IR-SED can also explain the observables above with a similarly high gas mass $\sim 2.1\times10^{-2}\,M_\odot$.
We confirm the conclusions of \citet{Ruaud2022} that optically thin \coo\ rotational lines provide reasonable estimates of the disk mass and can therefore be used as gas disk tracers.
\end{abstract}


\section{Introduction}
\label{sec:intro}

Disks of gas and dust around young stars (hereafter, protoplanetary disks) are the sites of planet formation, and their mass is fundamental to understanding when and how planets and small bodies form.
While the gas content sets limits on the potential masses of forming giant planets, the dust mass constrains the masses and formation times for the cores of gaseous planets and terrestrial planets.
The gas-to-dust mass ratio ($\Delta_\mathbf{gd}$), moreover, is an indicator of the relative rates of planet formation and gas disk dispersal and indicates the stage of disk evolution and planet formation (e.g., \citealt{Miotello_PPVII_2022} for a recent review).
Ideally, independent and reliable dust and gas mass estimations are needed to infer the disk physics and evolution, but measuring both masses is complicated and challenging.

Dust masses ($M_\mathrm{dust}$) are estimated by the dust thermal emission at (sub)millimeter wavelengths, which is sensitive to particles with sizes $\la 1\,\mathrm{cm}$ and is mostly optically thin (e.g., \citealt{Ansdell2016, Pascucci2016ApJ...831..125P}).
However, $M_\mathrm{dust}$ estimates rely on the dust opacity $\kappa_\nu$ which depends on the composition of dust grains and their size distribution.
Therefore, estimates of $M_\mathrm{dust}$ from a single flux measurement strongly depend on the assumptions made on the dust properties \citep[e.g.,][]{Miotello_PPVII_2022}.
Improved estimates of $M_\mathrm{dust}$ can be made by fitting the spectral energy distribution (SED) at long wavelengths ($\ga 100\,\micron$) where the emission is typically optically thin \citep[e.g.,][]{woitke_consistent_2019}.

Gas masses ($M_\mathrm{gas}$) are more difficult to estimate since there are very few optically thin gas emission lines that may trace the disk mass reservoir.
$\mathrm{H_2}$ is the most  abundant molecule in the gas phase in the disk, but its emission is faint.
This is because $\mathrm{H_2}$ is a light, homonuclear molecule with no permanent dipole moment and hence has only transitions at high energy levels ($E_u \sim$ few  100$-$1000K), while the majority of the gas in the disk around T-Tauri stars is far colder ($\sim 30\,\mathrm{K}$).
The less abundant isotopologue $\mathrm{HD}$ is favored to measure $M_\mathrm{gas}$, although it also traces relatively warm gas (needed to excite the first rotational level of $\mathrm{HD}$ at $E_u \sim 128\,\mathrm{K}$), and therefore has some limitations on its suitability as a mass tracer \citep{Trapman2017_HD, Ruaud2022}.

Carbon monoxide ($\mathrm{CO}$) is the most abundant molecule after $\mathrm{H_2}$ and is co-spatially distributed with $\mathrm{H_2}$ at the disk surface.
In the disk mid-plane, $\mathrm{CO}$ freezes out on the dust grain surface (when $T_\mathrm{dust} \lesssim 20\,\mathrm{K}$) where it can be processed into more refractory ices.
With its high detectability at (sub)millimeter wavelengths in disks, $\mathrm{CO}$ and its isotopologues have long been considered among the best tracers of gas disk mass.
However, recent Atacama Large Millimeter/submillimeter Array (ALMA) observations of Class-II disks have cast doubts about its ability as a mass tracer because model-predicted line emissions of $\mathrm{CO}$ and its isotopologues are higher than observed even after accounting for the fact that $\mathrm{CO}$ freezes-out in the mid-plane \citep[e.g.,][]{Ansdell2016, miotello_lupus_2017, Long2017ApJ...844...99L}.
This raises questions as to whether CO chemical abundances in disks differ from that in the interstellar medium (ISM) or whether the disk gas masses are low.
Furthermore, the $\mathrm{CO}$-based $M_\mathrm{gas}$ were smaller by $\sim 1 -2$ orders of magnitude compared with the $\mathrm{HD}$-based values for the few disks where $\mathrm{HD}$ has also been detected \citep[e.g.,][]{Bergin2013_HD_Natur.493..644B, McClure2016_HD_ApJ...831..167M, Trapman2017_HD}.
Thus, some works have argued for higher gas masses but large-scale depletion of $\mathrm{CO}$ due to dynamical processes that sequester $\mathrm{CO}$ into forming planetesimals and proto-planets \citep[e.g.,][]{BerginandWilliams2017, Bosman_Banzatti_2019A&A...632L..10B, Sturm2022A&A...660A.126S}. 

A different solution was proposed recently by \citet{Ruaud2022} (hereafter RGH22), who argued that by including (a) the density distribution given by self-consistent vertical hydrostatic pressure equilibrium, (b) isotopologue-selective chemistry, and (c) grain-surface chemistry where $\mathrm{CO}$ to $\mathrm{CO_2}$ conversion is a key reaction, the apparent discrepancy between the $\mathrm{HD}$ and $\mathrm{C^{18}O}$ derived masses can be resolved.
They concluded that CO chemistry in disks is in fact similar to that in the ISM and that the optically thin lines from $\mathrm{C^{18}O}$ can be used as a gas mass tracer.
Although they could retrieve typical $\mathrm{C^{18}O}$ fluxes observed for the Lupus sample, they did not consider individual disks in detail or compare the profile and radial distribution of the line emission.

In this work, we develop a tool to estimate the disk mass: \texttt{DiskMINT} (Disk Model for INdividual Targets).
It uses the dust temperature-based approach suggested in RGH22: generating a self-consistent gas disk structure on top of a SED-derived dust disk. 
It also uses a reduced chemical network that properly captures the conversion of $\mathrm{CO}$ into $\mathrm{CO_2}$ ice.
The tool is tested in considerable detail for the Class~II source RU~Lup. 
We select RU~Lup because this disk has been previously inferred \citep{miotello_lupus_2017} to have a low gas mass of $M_{\mathrm{gas}} \sim 1.5\times10^{-3}\,M_\odot$ with $\Delta_\mathrm{gd} \sim 4$  which is at odds with the large mass accretion rate onto the star \citep{alcala_x-shooter_2017} and the large disk size \citep{HuangDSHARP2018ApJ...869L..42H, HuangRULup2020ApJ...898..140H}.

The paper is organized as follows. 
First, we describe the modeling procedure in Section~\ref{sec:modeling}.
Then, we summarize the stellar parameters, observational data, and model setup for RU~Lup in Section~\ref{sec:apptoRULup}, followed by the results and discussion in Section~\ref{sec:result}.
We present our summary and outlook in Section~\ref{sec:summary}.

\section{Model Description}
\label{sec:modeling}

\texttt{DiskMINT} is a dust temperature-based disk model, and uses the recommendations made by RGH22.
From their analysis using a full thermo-chemical model that includes isotope-selective photodissociation and 3-phase grain-surface chemistry, RGH22 identified two main components that can be used to construct a simplified model to accurately simulate $\mathrm{C^{18}O}$ emission.
The two components are: (a) a self-consistent disk physical structure, based on the dust temperature $T_\mathrm{d}$ and imposing vertical hydrostatic pressure equilibrium (hereafter VHSE) to calculate densities consistent with this vertical temperature; and (b) a reduced chemical network that includes isotope-selective photodissociation and grain-surface chemistry that accounts for conversion of $\mathrm{CO}$ into $\mathrm{CO_2}$ ice (see Appendix~A of RGH22).
Since $\mathrm{C^{18}O}$ traces the vertical layer where gas temperature $T_\mathrm{g}$ is still very similar to $T_\mathrm{d}$, RGH22 found that a simplified dust disk structure model (which does not consider the self-consistent gas temperature $T_\mathrm{g}$ computed from full thermal equilibrium) can be used to estimate the $\mathrm{C^{18}O}$ emission.
As such, we build \texttt{DiskMINT} based on this simplified model.

\begin{figure}
  \gridline{\fig{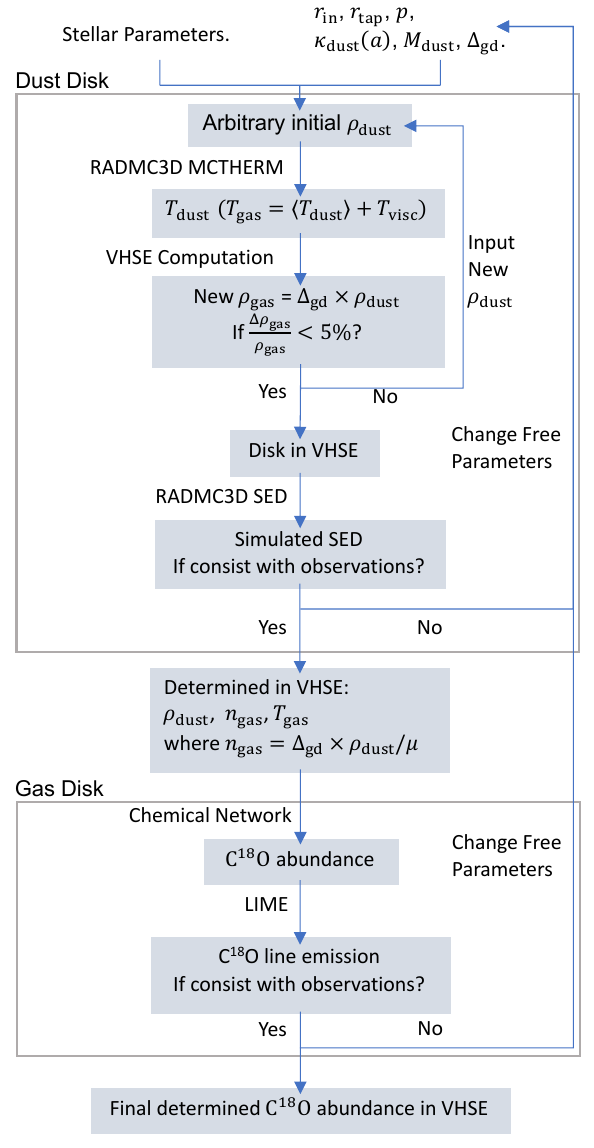}{0.47\textwidth}{}
            }
  \caption{
  Flow chart summarizing our modeling approach.
  Free parameters are listed at the top of the chart, and stellar parameters are fixed.
  These free parameters include the disk's inner radius ($r_\mathrm{in}$), tapering-off radius ($r_\mathrm{tap}$), power-law index of the surface density distribution ($p$), dust opacity ($\kappa_\mathrm{dust}(a)$), dust disk mass ($M_\mathrm{dust}$), and the gas-to-dust mass ratio ($\Delta_\mathrm{gd}$). In this model, we assume that the dust and gas are well-coupled. Therefore, the gas temperature ($T_\mathrm{gas}$) represents a cross-section weighted mean value of the dust grain temperature ($T_\mathrm{dust}$), adding the contribution from viscous heating ($T_\mathrm{visc}$). Similarly, the gas mass density ($\rho_\mathrm{gas}$) and number density ($n_\mathrm{gas}$) are derived from the dust density ($\rho_\mathrm{dust}$) by multiplying it with the factors of $\Delta_\mathrm{gd}$ and $\Delta_\mathrm{gd}/\mu$, respectively, where $\mu$ represents the mean molecular mass.
  }
  \label{fig1:flowchart}
\end{figure}

The overall method adopted in our analysis is summarized in the flow chart shown in Figure~\ref{fig1:flowchart}. 
Two main steps are involved in obtaining a self-consistent disk model that fits the $\mathrm{C^{18}O}$ line and continuum data.
The goal of Step~1 is to find a density structure --- based on the dust temperature profile --- that is self-consistent with pressure equilibrium and fits the SED. 
This is achieved by iteration: starting from an arbitrary initial density, computing the dust temperature using \texttt{RADMC-3D} \citep[Version 2.0,][]{Dullemond_radmc-3d_2012}, determining the resulting gas temperature, solving for vertical hydrostatic pressure equilibrium, and subsequently updating the density and temperatures in iterations until convergence.
Step~2 computes the $\mathrm{C^{18}O}$ abundance distribution via the reduced chemical network that includes isotopologue-selective dissociation and CO to CO$_2$ ice conversion on grains. 
It then computes the $\mathrm{C^{18}O}$ line emission using the radiative transfer tool \texttt{LIME} (Line Modelling Engine Version 1.9.5, \citealt{BrichHogerheijde_LIME_2010}) and compares it to the observed line emission profiles. 
If the agreement is poor, then the initial parameters (e.g., surface density distribution $\Sigma$, gas-to-dust mass ratio $\Delta_{\mathrm{gd}}$) are modified to repeat the entire modeling procedure from Step 1 until a satisfactory match with both SED and line emission is obtained. 
Details about the two steps are provided in the following subsections. 

\subsection{Model Step 1: Finding a Self-consistent Disk Structure that Fits the SED}
\label{subsec:MODELSTEP-1}

The main input parameters for this step (apart from the stellar parameters) are the surface density distribution, dust size distribution and opacity, and the disk gas-to-dust ratio. 
\paragraph{Surface Density}
The surface density distribution is assumed to be that of a viscously evolving disk \citep[e.g.,][]{Hartmann1998} and is specified as:
\small
\begin{equation}
    \Sigma(r) = \Sigma_1 (\frac{r}{1\,\mathrm{AU}})^{-p} \exp{\left[- (\frac{r}{r_{\mathrm{tap}}})^{2-\gamma}\right]};    (r_{\mathrm{in}} < r< r_{\mathrm{out}})
\end{equation}
\normalsize
where $r$ is the radial distance from the star, $\Sigma_1$ is the surface density at $1\,\mathrm{AU}$ that is scaled according to the chosen disk mass, $\gamma$ is the tapering-off exponent and $p$ is the power-law index.
We further assume that $\gamma = p$, which represents the self-similar viscous solution.
The inner radius cut-off, $r_{\mathrm{in}}$, is assumed to be the dust sublimation radius while the outer radius $r_{\mathrm{out}}$ is chosen to be much larger than $r_\mathrm{tap}$ to ensure that all the mass is included. 

\paragraph{Dust Properties} 
As discussed later, the gas temperature is computed assuming an equilibrium value from collisional heating/cooling by dust grains. 
To determine the gas temperature accurately, we use multiple dust sizes and calculate the size-dependent dust temperature. 
The dust species are divided into multiple grain-size bins equally distributed in log-space.
The dust number density follows a power-law distribution: 
$
    n(a) \propto a^{-q}\ \mathrm{with}\ a\in[a_\mathrm{min}, a_\mathrm{max}] 
$
where $a$ is the dust grain size and $q$ is the exponent describing the size distribution.
We adopt a dust composition consisting of 64\% astronomical silicates and 36\% graphite by volume fraction, which is representative of ISM dust with the ratio of visual extinction to reddening $R_V = 5.5$ \citep[][hereafter WD01]{WeingartnerDraine2001}.
A similar composition has been adopted in many previous disk models \citep[e.g.,][]{Ansdell2016, Miotello2016, woitke_consistent_2019}.
The \texttt{dsharp\_opac} package from \citet{Birnstiel_DSHARP_2018} is used to compute the wavelength dependence of dust opacity, and the optical constants are those of astrosilicate from WD01 and graphite from \citet{Draine2003_ApJ_II...598.1026D}. 

\paragraph{Gas-to-dust ratio}
In order to determine the vertical hydrostatic pressure equilibrium solution, the gas pressure gradient and hence the gas density are needed. 
In \texttt{DiskMINT}, the surface density distribution of gas and dust can in principle be specified separately as a function of radius, and this determines the local gas-to-dust ratio $\Delta_\mathrm{gd}(r)$. 
However, for our modeling of RU~Lup, we assumed a constant value throughout the disk for simplicity, which, as we show later, can already match the $\mathrm{C^{18}O}$ data.

The vertical dust density distribution, $\rho_{{\rm d}}(r, z)$, is initially set as an arbitrary Gaussian profile.
This is then distributed according to the mass fraction in each grain size bin to obtain $\rho_{{\rm d}}(r, z, a)$.
\texttt{RADMC-3D} \citep{Dullemond_radmc-3d_2012} is used to compute the dust temperature $T_\mathrm{d}(r, z, a)$ for each grain size bin.
We first determine the gas temperature $T_\mathrm{g}(r, z)$ balancing collisional energy exchange with dust grains; this contribution is denoted as $T_\mathrm{g,d}$. 
Since RU~Lup is a high accretor, the near and mid-infrared SED can be affected by viscous heating \citep{BossYorke1996ApJ_SED}. 
\texttt{RADMC-3D} does not currently include this viscous heating term, and we hence add this as a separate contribution to the gas ($T_\mathrm{g,v}$), and the dust as described later below. 

$T_\mathrm{g,d}$ is estimated from the following equation balancing dust heating and cooling

\begin{equation}\label{eq:Tgas1}
\begin{array}{cc}
\displaystyle\sum_{T_\mathrm{d}(a) > T_\mathrm{g,d}} & A_\mathrm{H} n_\mathrm{d}(a) \pi a^2 n_\mathrm{H} \bar{v}_\mathrm{H} 2k_B [T_\mathrm{d}(a) - T_\mathrm{g,d}] \\
= \displaystyle\sum_{T_\mathrm{d}(a) < T_\mathrm{g,d}} & A_\mathrm{H} n_\mathrm{d}(a) \pi a^2 n_\mathrm{H} \bar{v}_\mathrm{H} 2k_B [T_\mathrm{g,d} - T_\mathrm{d}(a)],
\end{array}
\end{equation}
where $T_\mathrm{d}(a)$ is the dust temperature at the grain size $a$, $A_\mathrm{H}$ is the mean accommodation coefficient, $n_\mathrm{d}(a)$ is the dust number density distribution, $n_\mathrm{H}$ is the gas number density, $\bar{v}_\mathrm{H}$ is the gas thermal velocity, and $k_B$ is the Boltzmann constant. 
This thermal balance equation simplifies to 
\begin{equation}\label{eq:Tgas2}
\begin{array}{cc}
\displaystyle\sum_{T_\mathrm{d}(a) > T_\mathrm{g,d}} n_\mathrm{d}(a) a^2\  [T_\mathrm{d}(a) - T_\mathrm{g,d}]\\
= \displaystyle\sum_{T_\mathrm{d}(a) < T_\mathrm{g,d}} n_\mathrm{d}(a) a^2 \ [T_\mathrm{g,d} - T_\mathrm{d}(a)],
\end{array}
\end{equation}
and only the terms related to dust size remain.

The gas temperature contributed by dust grain collisions ($T_\mathrm{g,d}$) is thus a cross-section weighted mean value between the hot (small) and cold (large) dust grain temperatures, and therefore the number of grain size bins ($N$) used could potentially affect the accuracy of the gas temperature evaluation.
We adopt $N=20$ as we find that this results in gas temperature deviations (caused by $N$) to be less than 5\%. 

We next estimate the temperature due to a balance between accretion heating and radiative cooling \citep[e.g.,][]{Armitage2022_LN_arXiv220107262A}.
Viscous heating is given by $(9/4)\nu\Sigma\Omega_k^2$ where $\nu$ is the kinematic viscosity, and $\Omega_K$ is the Keplerian angular frequency.
Cooling is given by $2\sigma_\mathrm{SB} T^4$, and for a disk accreting in steady state the accretion rate ($\dot{M}_\mathrm{acc}\sim 3\pi\nu\Sigma$) we have
\begin{equation}
    T_\mathrm{g,v} = \left[ \frac{3 G M_\star \dot{M}_\mathrm{acc}}{8 \pi \sigma_{\rm SB} r^3} \times ( 1 - \sqrt{\frac{r_\star}{r}}) \right]^{\frac{1}{4}},
\end{equation}
where $G$ is the gravitational constant and $\sigma_\mathrm{SB}$ is the Stefan–Boltzmann constant.
The resulting gas temperature is determined by adding the two temperatures in quadrature and is therefore given by 

\begin{equation}
    T_\mathrm{g}(r, z) = \left[T_\mathrm{g,d}^{4}(r, z) + T_\mathrm{g,v}^{4}(r, z)\right]^{\frac{1}{4}}.
\end{equation}
Viscous heating dominates only at the mid-plane in the inner disk ($\la 10\,\mathrm{AU}$) for typical disk densities \citep[also see, e.g.,][]{D'Alessio1998_viscous_ApJ...500..411D}.

Once the gas temperature is computed, the new density structure is calculated from the pressure gradient by solving
\begin{equation}\label{eq:VHSE}
 \frac{dP(r, z)}{dz} = -\rho_{\mathrm{gas}}(r, z) \Omega^2 z,
\end{equation}
where $P(r, z)$, $\rho_{\mathrm{gas}}$ and $\Omega$ are the gas pressure, gas density (assumed to be the total dust density times a constant $\Delta_\mathrm{gd}$) and Keplerian frequency, respectively. For the next iteration, the dust density profile with $z$ is rescaled with this vertical gas density profile, and re-normalized to the surface density at this radius.
The dust temperatures are re-calculated with the new dust density distribution using \texttt{RADMC-3D}. 
The steps above are recomputed until convergence is achieved at the iteration $m$: $| \left[\rho_{\mathrm{gas}, m}(r, z) - \rho_{\mathrm{gas}, m-1}(r, z) \right] /\rho_{\mathrm{gas}, m-1}(r, z) | < 5\%$ for regions with $\rho_\mathrm{gas} > 10^{-20}\,\mathrm{g\,cm^{-3}}$ (corresponding to $n_{\rm H} \ga 10^{3}\,\mathrm{cm^{-3}}$). 
The error tolerance was chosen as a reasonable compromise between accuracy and speed of computation ($\sim 4$ hours to achieve convergence when running with 24 threads with 2.10\,GHz CPUs).
Lower tolerances did not significantly change the results.

The above procedure results in a dust and gas density and temperature distribution which are all self-consistent with the local vertical pressure gradient.
We described viscous heating for gas above, but this term is also relevant for heating dust grains.
Since this is difficult to incorporate into the \texttt{RADMC-3D} code, we include this effect by adding it to the dust grains before computing the SED.
This is done by considering the gas as a thermal reservoir that equilibrates the dust temperature in regions where dust and gas are highly coupled.
In practice, we estimate the extent of this mid-plane region as the region where the temperature differences between the hottest/smallest grain and coldest/largest grain are small enough as $|\left(T_\mathrm{d}(a_\mathrm{min}) - T_\mathrm{d}(a_\mathrm{max})\right)/T_\mathrm{d}(a_\mathrm{max})| < 10\%$. $T_\mathrm{d}(a) = T_\mathrm{g}$ is set for all grain sizes in this coupled region.
We then run \texttt{RADMC-3D} to compute the SED and compare it with the observed SED. 

We vary the disk dust parameters until a satisfactory match to the SED is obtained.
The dust opacity $\kappa_\nu$ and the dust mass $M_\mathrm{dust}$ are two main parameters affecting the synthetic SED: Changing $\kappa_\nu$ alters the slope of the long-wavelength portion of the SED and $M_\mathrm{dust}$ moves the flux density up and down.
In practice, we find the best fit $\kappa_\nu$ by comparing the slope of the dust opacity $\beta_\mathrm{abs} = - \frac{d \log(\kappa_{\mathrm{abs}})}{d \log{\lambda}}$ with the slope of the SED at long wavelength $\alpha_\mathrm{SED} = - \frac{d \log{F_\nu}}{d \log{\lambda}}$ ($\lambda \gg 100\,\micron$) based on the relation between the two slopes $\beta_\mathrm{abs} = \alpha_\mathrm{SED} - 2$.
When the dust composition is fixed, we first vary the maximum particle size $a_\mathrm{max}$ and keep the slope of the number density distribution with size fixed to $q = 3.5$, which is the value expected in collisional equilibrium \citep{Birnstiel2011_dustsize_A&A...525A..11B}.
If the upper limit of $a_\mathrm{max} = 1\,\mathrm{cm}$ is reached while varying $a_\mathrm{max}$, then $q$ is varied to find the best match of the slope.
After the best-fit $\kappa_\nu$ is found, the $M_\mathrm{dust}$ is derived by matching the absolute value of the flux density at long wavelengths.

\subsection{Model Step 2: Computing the \texorpdfstring{$\mathrm{C^{18}O}$}\ \ Line Emission and Profile}
\label{subsec:MODELSTEP-2}

The next step in our modeling approach is to run the reduced chemical network described in RGH22 to obtain the $\mathrm{C^{18}O}$ abundance with $(r,z)$.
The photodissociation rates (for our application target RU~Lup) are computed from the UV \emph{HST}/COS median-resolution spectrum obtained by \citet{France+2014_UVSpec} (see also Figure~\ref{fig2:observations} for average photometric values from this spectrum).
We assume all gas is molecular in the disk structure calculation but explicitly solve for the chemistry by specifying the corresponding H nuclei density ($n_\mathrm{H}$) for the chemical network. 
This means that all molecular abundances in the chemical network are defined by their density ratio compared to the density of H nuclei.
Finally, the gaseous abundances of $\mathrm{C^{18}O}$ and the disk structure are inputs to \texttt{LIME} \citep{BrichHogerheijde_LIME_2010} to compute the non-LTE(local thermal equilibrium) synthetic $\mathrm{C^{18}O}$ (2-1) and (3-2) emission.

The model parameters are varied until the synthesized SED and $\mathrm{C^{18}O}$ line emission match the observations. We fix $\kappa_\nu$ and $M_\mathrm{dust}$ to the values determined in the SED fitting, and explore a range of gas-to-dust ratios $\Delta_\mathrm{gd} = 5, 10, 50, 100$ (which covers the low disk $\Delta_\mathrm{gd}$ reported in the literature up to the ISM value) to generate a grid of $M_\mathrm{gas}$. Since the self-consistent VHSE solution depends on the gas mass (which varies with $\Delta_\mathrm{gd}$ in the gas mass grid), the vertical density structure of each of these models slightly differ.
However, the SED at long wavelengths traces the optically thin thermal emission from the large grains and remains the same as it is not sensitive to the vertical dust density distribution. The derived $M_\mathrm{dust}$ therefore remains unaltered even as $\Delta_\mathrm{gd}$ is varied. 
We start from the beginning for each grid point and find that we do not need to re-fit the SED, hence we calculate the dust thermal structure with \texttt{RADMC-3D}, solve the VHSE through iterations, and then derive the \coo\ abundance by the reduced chemical network.
Next, the $\mathrm{C^{18}O}$ line luminosity ($L_\mathrm{C^{18}O}$) is computed to compile a $L_\mathrm{C^{18}O}$ vs. $M_\mathrm{gas}$ relation.
The best-fit $M_\mathrm{gas}$ is then determined as the value where the modeling relation ($L_\mathrm{C^{18}O}$ vs. $M_\mathrm{gas}$) intersects the luminosity inferred from the observations.  
Finally, we run the model with best-fit $M_\mathrm{gas}$ again also from the beginning to verify the estimate found above.

In this work, we not only compare total line luminosities as in RGH22 but also match the velocity profile and radial distribution of the $\mathrm{C^{18}O}$ (2-1) line.
These are generated by the \texttt{Python} package \texttt{GoFish} \citep{Teague2019_GoFish_JOSS....4.1632T} from the simulated \texttt{LIME} image and follow the same procedure used on observational data.
The slope of the surface density distribution and the gas-to-dust ratio as a function of radius are parameters that can be changed to improve the fit on the line profile, if necessary.


\section{Application to RU~Lup}
\label{sec:apptoRULup}


\subsection{The Highly Accreting RU~Lup Star and Its Dust and Gas disk}
\label{subsec:obs}

\begin{figure*}
  \gridline{\fig{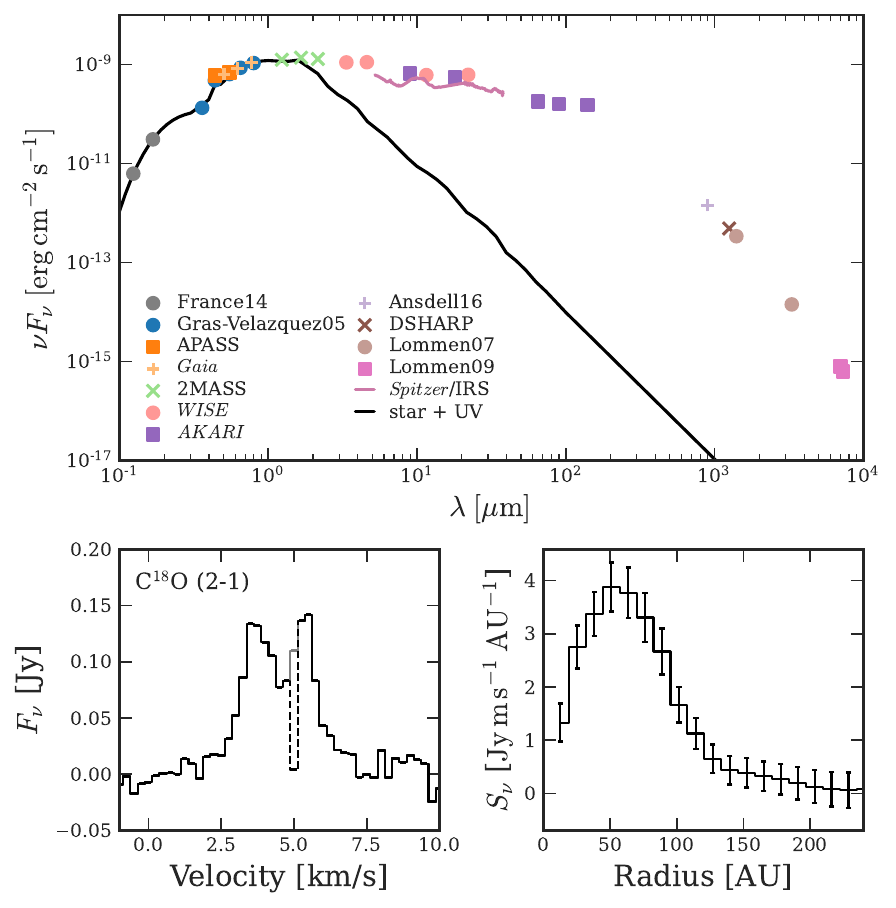}{0.60\textwidth}{}
            }
  \caption{
    Observational data that used in this work.
    Top panel: spectral energy distribution (SED). Photometry in colored markers and \textit{Spitzer/}IRS spectrum in magenta. 
    A combined \texttt{BT-settl} stellar photosphere spectrum \citep{Allard_2003IAUS..211..325A, Allard_2011ASPC..448...91A} with a 10,500 K blackbody radiation fitting the UV data-points is shown as a black line. 
    Lower panels: ALMA $\mathrm{C^{18}O}$ (2-1) velocity profile (lower left) and radial profile (lower right), data from \citet{HuangRULup2020ApJ...898..140H} with uncertainties calculated by \texttt{GoFish}.
    Linear interpolation (grey line) is adopted to recover the channel contaminated by a dark cloud (dashed black line) in the line spectrum.
    SED references: photometry from the \emph{HST} spectrum in \citet{France+2014_UVSpec}; optical photometry from \citet{Gras-V2005A&A...443..541G}, APASS \citep{Henden2016_APASS_yCat.2336....0H}, \emph{Gaia} \citep{gaia_collaboration_gaia_2018}; infrared photometry from 2MASS \citep{Skrutskie2006_2MASS_AJ....131.1163S}, \emph{WISE} \citep{Wright2010_WISE_AJ....140.1868W}, \emph{AKARI} \citep{Ishihara2010_AKARI_A&A...514A...1I}; and (sub)millimeter data from \citet{Ansdell2016}, DSHARP \citep{andrews_disk_2018}, and \citet{Lommen2007A&A...462..211L, Lommen2009A&A...495..869L}.
  }
  \label{fig2:observations}
\end{figure*}

RU~Lup (Sz~83, 2MASS~J15564230-3749154) is a K7-type star located at a distance of $158.9\,\mathrm{pc}$ \citep{gaia_collaboration_gaia_2018} and a member of the Lupus~II star-forming region \citep{Comeron2008_handbook_hsf2.book..295C}.
RU~Lup has the highest mass accretion rate ($\sim 10^{-7}\, M_\odot$/yr, \citealt{alcala_x-shooter_2017}) and is one of the most active stars in the region with large irregular variations in both spectroscopy and photometry from ultraviolet (UV) to infrared (IR) wavelengths \citep[e.g.,][]{hughes_stellar_1994, Herczeg2005AJ....129.2777H, Gahm2013A&A...560A..57G}.
The stellar mass estimates range from 0.2 to 1.2\,$M_\odot$ \citep[e.g.,][]{alcala_x-shooter_2017,  AndrewsScaling2018ApJ...865..157A,YenStellarMass2018A&A...616A.100Y}. 
Here, we adopt the value of $0.7\,M_\odot$ from more recent evolutionary models \citep{alcala_x-shooter_2017} over the dynamical mass of 0.2\,$M_\odot$.
This is because the disk of RU~Lup is close to face-on which introduces a large uncertainty in the dynamical mass \citep{YenStellarMass2018A&A...616A.100Y}.
As one of the most extensively observed Class~II objects in Lupus, photometry and spectra are available from the UV to radio wavelengths resulting in the multi-wavelength spectral energy distribution (SED) shown in Figure~\ref{fig2:observations}, where average photometry is reported for multi-epoch observations. 

A large-scale, complex proto-planetary disk has also been recently revealed by ALMA.
The millimeter dust disk appears symmetric with multiple annular gaps and rings and extends out to a radius of $\sim 63\,\mathrm{AU}$ \citep{HuangDSHARP2018ApJ...869L..42H, HuangRULup2020ApJ...898..140H}. 
In contrast, CO emission has a more asymmetric morphology. 
\citet{HuangRULup2020ApJ...898..140H} identified a Keplerian disk with a radius of $\sim 120\,\mathrm{AU}$, similar in size to that inferred via scattered light \citep{Avenhaus2018}, surrounded by an envelope extending out to $\sim 260\,\mathrm{AU}$ with spiral arms and clumps.

However, the $\mathrm{C^{18}O}$ emission, which we focus on and aim to model in this work, is less complex.
The $\mathrm{C^{18}O}$ emission is symmetric, only traces the Keplerian disk, and has a radius of $\lesssim 100\,\mathrm{AU}$.
The lower panels of Figure~\ref{fig2:observations} show the \coo\ (2-1) line profile and radial intensity cut from publicly available datacubes \citep{HuangRULup2020ApJ...898..140H} generated using \texttt{GoFish}. 
We choose the same aperture and wavelength range used in \citet{HuangRULup2020ApJ...898..140H}, $1.5\arcsec \sim 240\,\mathrm{AU}$ and $1.75-7.25\,\mathrm{km/s}$, as the maximum extent to include all the emitting areas and channels when computing the line profile and radial profile.
In the line profile, there is clear dark cloud contamination at $5\,\mathrm{km/s}$ (dashed line).
Linear interpolation (grey point and line) is utilized to recover the disk emission in this channel, which brings the integrated total flux of \coo\ (2-1) from $0.34 \pm 0.03\,\mathrm{Jy\,km/s}$ to $0.37 \pm 0.03\,\mathrm{Jy\,km/s}$.
For the radial profile, deprojection is applied using the disk position angle $\mathrm{PA} = 121\arcdeg$ and inclination $i = 18.8\arcdeg$ \citep{HuangDSHARP2018ApJ...869L..42H}.


The dust and gas mass of RU~Lup, hence the gas-to-dust mass ratio $\Delta_{\mathrm{gd}}$, have been previously estimated  using continuum millimeter emission and CO isotopologue emission.
\citet{Ansdell2016} measured the $\mathrm{{}^{13}CO}$(3-2) and $\mathrm{C^{18}O}$(3-2) line fluxes and compared them to a grid of simple disk models by \citet{WilliamsBest2014}: They inferred a gas disk mass of $\sim 2.7_{-1.7}^{+7.3}\times10^{-3} \,M_\odot$ and a gas-to-dust mass ratio $\Delta_\mathrm{gd} \sim 8.9_{-5.7}^{+24.4}$. 
\citet{Miotello2014, Miotello2016, miotello_lupus_2017} included isotope-selective dissociation in the thermo-chemical physical code \texttt{DALI} \citep{Bruderer_DALI_2012, Bruderer_DALI_2013} and used the same line luminosities to infer an even lower gas disk mass ($\sim 1.5_{-1.0}^{+2.5}\times10^{-3} \,M_\odot$) and gas-to-dust ratio ($\Delta_\mathrm{gd} \sim 3.8_{-2.6}^{+6.2}$).
Clearly, the low inferred disk mass and gas-to-dust mass ratio are hard to reconcile with the large dust and gas disk of RU~Lup and the high accretion rate onto the star; we therefore re-examine the dust and gas mass constraints using the \texttt{DiskMINT} modeling approach.

\subsection{Specific Models}
\label{subsec:specificmodels}

Two models are considered in this work with different vertical density distributions: (a) the VHSE model uses a self-consistent vertical hydrostatic pressure equilibrium solution; (b) the Gaussian model uses a parameterized Gaussian vertical structure.
Both models share the same surface density distribution and use the same dust grains (same $\kappa_\nu$ and $M_\mathrm{dust}$) determined by fitting the long wavelength portion of the SED ($\lambda \ga 100\,\micron$). 
The Gaussian model additionally fits the IR wavelengths ($10\,\micron \la \lambda \la 100\,\micron$) by assuming the pressure gradient to be a free parameter and thus varying pressure scale height in the Gaussian structure: $H_p = H_{p, \mathrm{100}}(r/100\,\mathrm{AU})^{\alpha}$ with free characteristic height $H_{p, 100}$ and flaring index $\alpha$.
This is the approach taken in a few recent studies to estimate disk masses and $\Delta_\mathrm{gd}$ \citep[e.g.,][]{woitke_consistent_2019, Zhang_MAPS_2021}.

The model input parameters are presented in Table~\ref{Tab:modelparameters}.
Stellar mass $M_\star$, radius $r_\star$ as well as mass accretion rate $\dot{M}_\mathrm{acc}$ are fixed and taken from the literature (see Section~\ref{subsec:obs}). 
The inner radius $r_\mathrm{in}$ is fixed at the dust sublimation radius, and the tapering-off radius is set as the dust outer radius $r_\mathrm{tap} \sim r_\mathrm{dust}$ given in \citet{HuangDSHARP2018ApJ...869L..42H}.
The dust opacity $\kappa_\nu$ is computed by \texttt{dsharp\_opac}: It uses a dust composition described in Section~\ref{subsec:MODELSTEP-1}, and has fixed volume fraction from $a_{\rm min} \sim 1.0\times10^{-6}\,\mathrm{cm}$ through to $a_{\rm max}$, in which $a_{\rm max}$ and the power law index $q$ are free parameters.
The other two free parameters are the dust disk mass $M_\mathrm{dust}$, and the gas-to-dust mass ratio $\Delta_\mathrm{gd}$.

The synthetic imaging setup for the models is obtained from observations (summarized in Section~\ref{subsec:obs}).
The output synthetic image is created with a pixel size of 0.04\arcsec, and with source distance, $i$ and $\mathrm{PA}$.
The image has $151 \times 151$ pixels to include all disk emission within 3.0\arcsec.
The dust continuum emission is also included in the synthetic image, and then the continuum is subtracted in the final line imaging datacube.
Then, the \texttt{LIME} output image is convolved with a beam of $0.32 \times 0.32 \,\arcsec$ to get the final synthesized image.

\begin{deluxetable*}{lcccc}
\tablecaption{Model Parameters\label{Tab:modelparameters}}
\tablewidth{0.9\textwidth}
\tablehead{
\colhead{Parameter} & \multicolumn{2}{c}{Symbol} & \multicolumn{2}{c}{Value}
}
\startdata
    \textit{Dust Properties} &&&& \\
    Volume fraction  &&& 64\% Silicate & 36\% Graphite \\
    minimum size & \multicolumn{2}{c}{$a_\mathrm{min}$}  &  \multicolumn{2}{c}{$1\times10^{-6}$ cm} \\
    maximum size & \multicolumn{2}{c}{$a_\mathrm{max}$}  &  \multicolumn{2}{c}{free parameter} \\
    exponential slope & \multicolumn{2}{c}{$q$}  &  \multicolumn{2}{c}{free parameter} \\ 
    \hline
    \textit{Radial Structure} &&&& \\
    inner radius of the disk & \multicolumn{2}{c}{$r_\mathrm{in}$}  &  \multicolumn{2}{c}{0.035 AU}\\
    tapering-off radius & \multicolumn{2}{c}{$r_\mathrm{tap}$}  &  \multicolumn{2}{c}{63 AU} \\
    surface density slope & \multicolumn{2}{c}{$p$} & \multicolumn{2}{c}{1} \\
    \hline
    \textit{Vertical Structure} &&& VHSE & Gaussian \\
    Characteristic Scale Height & \multicolumn{2}{c}{$H_{p, \mathrm{100}}$} & solved & free parameter\\
    Flaring Index & \multicolumn{2}{c}{$\alpha$} & solved & free parameter
    \\
\enddata
\tablecomments{In principle, all parameters in this table could be varied to fit the observations. 
However, only $a_\mathrm{max}$ and $q$ are changed here for the VHSE model as the default settings for other parameters could already give a good fit.
$H_{p, \mathrm{100}}$ and $\alpha$ are also set free for the Gaussian model while the vertical structure for the VHSE model is solved self-consistently from pressure equilibrium.
The best-fit free parameters are summarized in Table~\ref{Tab:modelresults}.}
\end{deluxetable*}

\section{Results and Discussion}
\label{sec:result}

The inferred dust parameters, dust and gas masses are summarized in Table~\ref{Tab:modelresults}. 
One of the main results of this work is that our model can explain RU~Lup's long-wavelength ($\lambda\gtrsim100\micron$) SED, the $\mathrm{C^{18}O}$ (2-1) and (3-2) line luminosities, and the velocity and radial profiles, with a higher $M_{\mathrm{gas}}$ and thus higher $\Delta_\mathrm{gd}$ than previously inferred.
We present details on these models in Section~\ref{subsec:modelresults}.
Effects of $\mathrm{CO} \leftrightarrow \mathrm{CO_2}$ conversion on grain-surface and differences between the Gaussian and VHSE models are discussed in Section~\ref{subsec:CompareLiterature}.
Our VHSE model under-estimates the strong IR excess of RU~Lup by a factor of $\lesssim 3$, and we discuss possible reconciliations in Section~\ref{subsec:discussIRSED}.

\begin{deluxetable*}{lccccccc}
\tablecaption{Model Main Results  \label{Tab:modelresults}}
\tablewidth{0.45\textwidth}
\tablehead{
\colhead{Model} & \colhead{$H_{p, \mathrm{100}}$} & \colhead{$\alpha$} & 
\colhead{$a_\mathrm{max}$} & \colhead{$q$} &
\colhead{$M_\mathrm{dust}$} & \colhead{$M_\mathrm{gas}$} &
\colhead{$\Delta_\mathrm{gd}$} \\
\colhead{} & \colhead{(AU)} & \colhead{} & 
\colhead{(cm)} & \colhead{} 
& \colhead{$(M_\odot)$} & \colhead{$(M_\odot)$} 
& \colhead{}
}
\startdata
    VHSE     & -  & -   & 0.3 & 3.5 &  $4.0\times 10^{-4}$ & $1.2\times 10^{-2}$ & 30  \\
    Gaussian & 30 & 1.1 & 0.3 & 3.5 &  $4.0\times 10^{-4}$ & $2.1\times 10^{-2}$ & 52 \\
\enddata
\end{deluxetable*}

\subsection{\texorpdfstring{\coo}\ \ Emission Indicates a Relatively High Gas Disk Mass for RU~Lup}
\label{subsec:modelresults}

\begin{figure*}
  \gridline{\fig{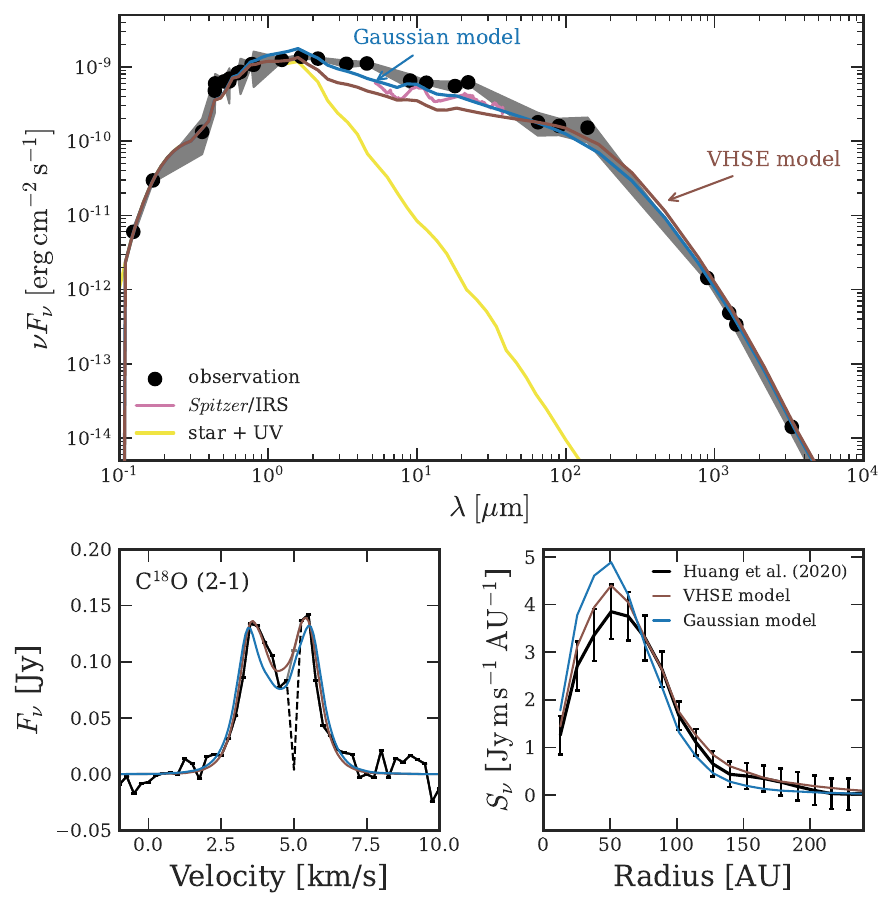}{0.60\textwidth}{}
            }
  \caption{
  Top panel: observed photometry compared to modeled SEDs.
  The uncertainties of the observation are shown in the gray shade. 
  The VHSE and Gaussian models are shown with brown and blue lines, respectively.
  Lower panels:$\mathrm{C^{18}O}$ (2-1) velocity profile (left) and radial cut (right) in black compared to modeled profiles. Color-coding for the models as in the top panel. 
  }
  \label{fig3:SEDandLineFitting}
\end{figure*}

In \texttt{DiskMINT}, the disk density structure is based on the dust temperature profile, and the dust disk is constructed by fitting the SED (See Section~\ref{subsec:MODELSTEP-1}).
The SED fits of the two models (VHSE and Gaussian) introduced in Section~\ref{subsec:specificmodels} are shown in the top panel of Figure~\ref{fig3:SEDandLineFitting}.
Both models share the same dust grain properties described in Table~\ref{Tab:modelparameters} and the best-fit free parameters are reported in Table~\ref{Tab:modelresults}.
The best-fit maximum grain size $a_\mathrm{max}$, $q$ parameters and dust disk mass $M_\mathrm{dust}$ are the same for both models: $a_\mathrm{max} = 0.3\,\mathrm{cm}$, $q = 3.5$ and $M_\mathrm{dust} \sim 4.0\times10^{-4}\,M_\odot$. 
Since the pressure scale height is determined using free parameters to match the SED in the Gaussian model, this model provides a better fit to the IR SED.
To find the best parameters, we start from the best-fit pressure scale height with $H_{p, 100} = 20.9324\,\mathrm{AU}$ and $\alpha = 1.1301$ reported in \citet{woitke_consistent_2019} for the disk of RU~Lup, and generate a grid of $H_{p, 100} = 15, 20, 25, 30\,\mathrm{AU}$ and $\alpha = 1.05, 1.10, 1.15, 1.20$.
Although we use a different dust composition and updated parameters for the central star, we find a relatively close pressure scale height with $H_{p, 100} = 30\,\mathrm{AU}$ and $\alpha = 1.10$ and a very similar synthetic SED for the Gaussian model.
For the VHSE models, the procedure of iteration to determine the vertical density structure to be consistent with the temperature profile sets the pressure scale height; there is no simple power law to describe the scale height thus the parameters $H_{p, 100}$ and $\alpha$ are not valid.

\begin{figure*}
    \gridline{\fig{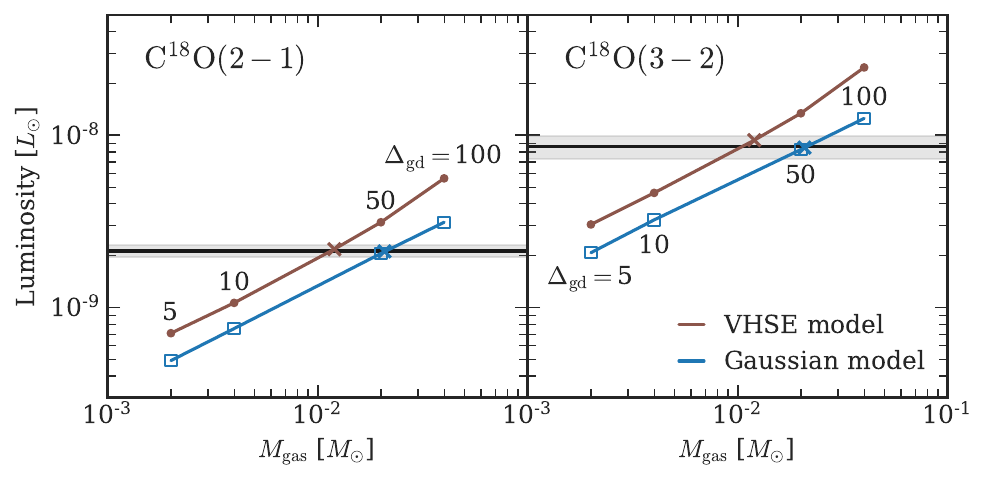}{0.70\textwidth}{}
            }
  \caption{
  Synthetic $\mathrm{C^{18}O}$ (2-1) and (3-2) luminosity vs. $M_\mathrm{gas}$.
  In both panels, the black horizontal line marks the luminosity inferred from the data and grey shades show their uncertainties: (2-1) line from \citet{HuangRULup2020ApJ...898..140H} and (3-2) from \citet{Ansdell2016}.
  The VHSE models introduced in this work are shown in points and line in brown.
  The Gaussian models are presented in squares and line in blue.
  In each modeling scenario in this work, four models with $\Delta_\mathrm{gd} = 5, 10, 50, 100$ were simulated, and one additional best-fit model was simulated at the cross-match `$\times$' point with observations.
  }
  \label{fig4:LM-C18O-2-1-method}
\end{figure*}

We then run the reduced chemical network and \texttt{LIME} to obtain the synthetic \coo\ luminosity, which is compared with the observation to obtain the best-fit $\Delta_\mathrm{gd}$ and hence the $M_\mathrm{gas}$ (see Section~\ref{subsec:MODELSTEP-2}).
The synthetic \coo\ (2-1) and (3-2) luminosities vs. $M_\mathrm{gas}$ using different $\Delta_\mathrm{gd}$ are presented in Figure~\ref{fig4:LM-C18O-2-1-method}. 
There are four data points on each modeling line representing $\Delta_\mathrm{gd} = 5, 10, 50, 100$ (points in Figure~\ref{fig4:LM-C18O-2-1-method}), and one additional best-fit model (`$\times$' in Figure~\ref{fig4:LM-C18O-2-1-method}), which is obtained at the cross point with the observation of $\mathrm{C^{18}O}$ (2-1) luminosity at left panels.
The best-fit gas masses for both models are within a factor of two: $M_\mathrm{gas} \sim 1.2\times10^{-2}\,M_\odot$ and $\sim 2.1\times10^{-2}\,M_\odot$ for the VHSE and Gaussian models, respectively.

In addition to matching the luminosity, the line spectrum and radial distribution from the synthetic \coo\ line images are also compared with the observations.
The lower panels of Figure~\ref{fig3:SEDandLineFitting} present the $\mathrm{C^{18}O}$\,(2-1) spectra and radial distribution for different models generated from simulated \texttt{LIME} datacubes with \texttt{GoFish} and the same setup (see Section~\ref{subsec:specificmodels}) as for the observational data.
These panels demonstrate that the VHSE model with default input parameters (Table~\ref{Tab:modelparameters}) also matches the $\mathrm{C^{18}O}$ (2-1) line profile and radial cut.
The Gaussian model can reproduce the line luminosity and matches the \coo\  (2-1) line velocity profile relatively well, but its emission is more compact than the VHSE model with the intensity peaking closer to the host star.
We note that the models also fit the \coo\ (3-2) luminosity \citep{Ansdell2016}, as shown in the right panel of Figure~\ref{fig4:LM-C18O-2-1-method}.
The \coo\ (3-2) line emission has a similar velocity profile and radial cut, but it is a factor of $\sim 5$ more luminous than the (2-1) line. For both models, even better fits may be achieved by changing the surface density distribution and by including a radial-dependent gas-to-dust ratio, but we did not consider these modifications necessary for RU~Lup.

\begin{figure*}
    \gridline{\fig{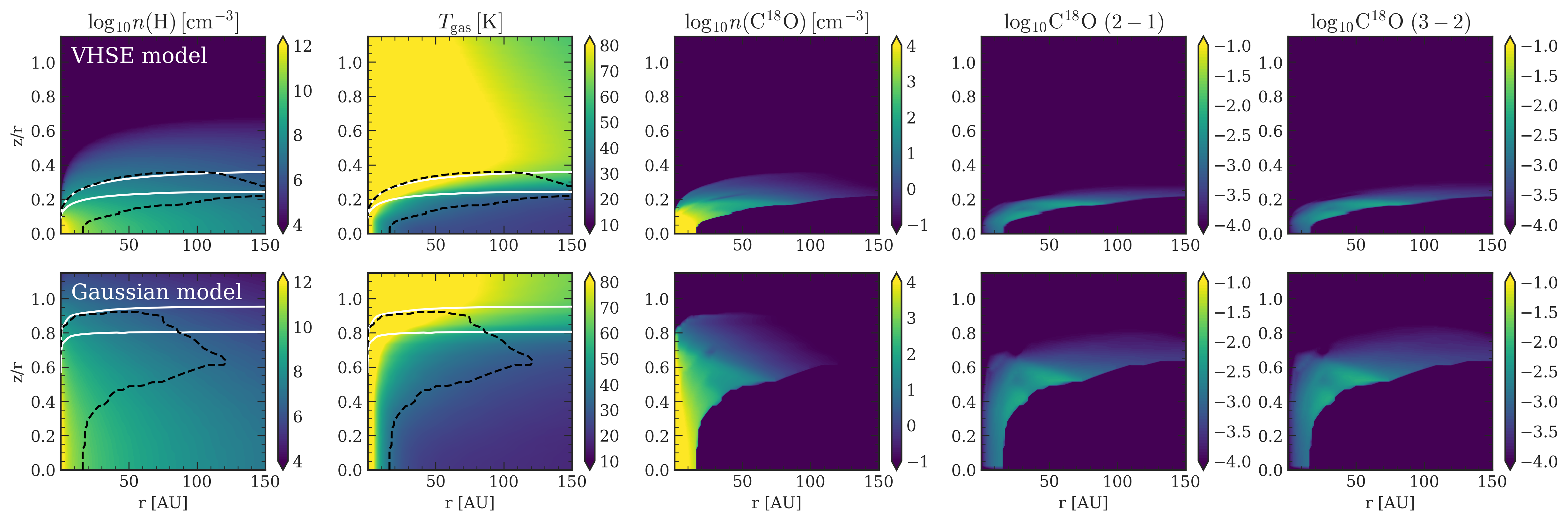}{0.98\textwidth}{}
            }
  \caption{Gas density distribution, gas temperature, $\mathrm{C^{18}O}$ density distribution, $\mathrm{C^{18}O}$ (2-1) and (3-2) line emitting layers (normalized to the total luminosity) are presented from left to right, VHSE model top and Gaussian model bottom row. In the first two panels, dashed black lines mark the boundaries of the C$^{18}$O location where $n(\mathrm{C^{18}O}) \ge 0.1$ and emitting region, and upper \& lower white lines represent the $A_{\rm V} \sim 1\,\&\,10$, respectively. While the $\mathrm{C^{18}O}$\,(3-2) flux is higher than the (2-1), the emitting layers are very similar.
  }
  \label{fig5:Dist-C18O}
\end{figure*}

The Gaussian model has an emission profile that is less radially extended compared with the observations.
This is because it has a very puffed-up density distribution which appears necessary to fit the IR SED: $H_{p, 100} = 30\,\mathrm{AU}$ by this work (also $H_{p, 100} \sim 21\,\mathrm{AU}$ reported in \citet{woitke_consistent_2019} which is a better match to the SED).
Since the scale height is parameterized as a power-law, this implies that the flaring index in the outer \coo\ emitting regions of the model disk is also higher.
The increased flaring moves the \coo\ emitting layer closer to the star and higher.
It is nearly a factor of $\sim 3-4$ higher than the VHSE disk at $r \sim 50\,\mathrm{AU}$ where most of the \coo\ emission comes from (Figure~\ref{fig5:Dist-C18O}). 
Although it is hard to obtain the height of the  \coo\ emitting layer for the RU~Lup disk due to its small inclination angle, this unrealistically puffed-up disk scenario -- with the emitting layer as high as $z/r \sim 0.8$ at $r \sim 50\,\mathrm{AU}$ -- is at odds with recent observations which instead find the \coo\ emitting layer of Class II disks to be at $z/r \sim 0.1$ for $r < 100\,\mathrm{AU}$ \citep{Paneque-Carreno_2023A&A...669A.126P}.
We note that, in principle, if the height of the emitting layer could be measured as it has been in some disks, then this information could be used to fit the \coo\ radial and velocity profiles for the Gaussian model.
We also find that if we assume the scale height obtained from the VHSE model and repeat the Gaussian modeling for RU~Lup, it results in a combination of ($M_{\mathrm{gas}}$, $L_{\mathrm{C^{18}O}}$) similar to the best-fit VHSE model, although the synthetic IR SED is no longer an improved match to the data.

While it may be possible to fit all of the observational data using a Gaussian disk model, determining the emission scale height requires very high spatial resolution observations and only works for disks with favorable inclination angles.
In their absence, the disk structure parameterization can deviate substantially from reality as we show for RU~Lup.
On the other hand, the VHSE model is physically motivated, determines the scale height at each radius via coupling of the disk density and temperature structure, and can simultaneously fit the radial and velocity distribution of flux.
Hence, we believe it is a more reliable indicator of conditions in the disk. 

In summary, our VHSE model fits the SED, total \coo\ line emission, velocity, and radial profiles from recent observations \citep[e.g.,][]{Ansdell2016, HuangRULup2020ApJ...898..140H} with relatively high $M_\mathrm{gas}$ and $\Delta_\mathrm{gd}$ ($\sim 30$) in comparison with the previously inferred $\Delta_\mathrm{gd}$ of $\sim 4$.
Using a Gaussian vertical distribution, our model also derives  a similarly high $M_\mathrm{gas}$ within a factor of $\sim 2$ of the one obtained from the VHSE model.
Thus, we conclude that the RU~Lup disk is not significantly low in its gas mass and nor has it undergone any substantial change in CO chemistry due to changes in C/H and O/H caused by planet formation processes.
We confirm the conclusions of RGH22, and find that optically thin \coo\ lines provide reasonable estimates of the disk mass. 
We also note that the RGH22 models compare favorably not only with the \coo\ fluxes, but also with the $\mathrm{{}^{13}CO}$, $\mathrm{CO}$, and atomic carbon forbidden line $\mathrm{[CI]}$ fluxes for a sample of large disks (R$\ge$200\,AU) \citep{Pascucci_2023arXiv230702704P}, and cold water emission as well (Ruaud \& Gorti, submitted).

\begin{figure*}
    \gridline{\fig{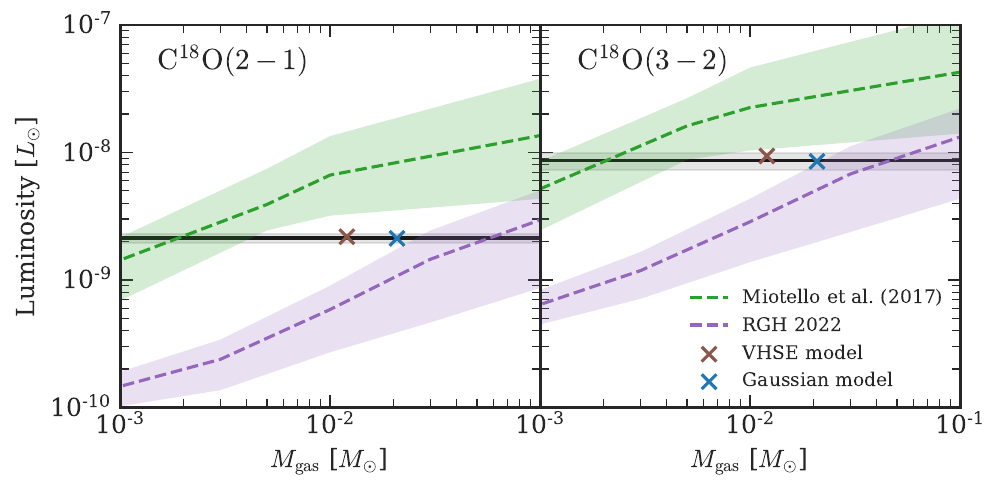}{0.70\textwidth}{}
            }
  \caption{
  Synthetic $\mathrm{C^{18}O}$ (2-1) and (3-2) luminosity vs. $M_\mathrm{gas}$.
  The observation results are presented in black.
  Two best-fit models are simulated at the cross-match point with observations represented by `$\times$' in brown and blue following the color in Figure~\ref{fig4:LM-C18O-2-1-method}.
  All possible results from full VHSE models in \citet{Ruaud2022} (RGH22) are shown in magenta shades.
  The Gaussian models without grain-surface chemistry from \citet{miotello_lupus_2017} are shown in green shades.
  The median value of all possible estimated luminosity for each $M_\mathrm{gas}$ is presented in dashed magenta and green lines, respectively.
  Our model predicts a relatively large gas disk mass ($\sim 1\times10^{-2}\,M_\odot$) lines between the model grid of RGH22 and \citet{miotello_lupus_2017}.
  }
  \label{fig6:LM-C18O-2-1}
\end{figure*}


\subsection{Comparisons with Literature Values}
\label{subsec:CompareLiterature}

Our work is the first to focus on specifically modeling CO isotopologue emission from RU~Lup, and matches the SED, \coo\ line luminosity, spectrum and radial profile.
In this section, we compare our source-specific model with the grids generated in previous works and discuss possible explanations for the different results in gas masses and gas-to-dust ratios.

First, we comment on the differences between the VHSE model presented here and those in RGH22.
Here our dust-temperature based model gives a factor of $\sim 2$ larger $L_{\mathrm{C^{18}O}}$ compared with the full VHSE thermo-chemical model by RGH22 (all possible results are shown in Figure~\ref{fig6:LM-C18O-2-1} magenta regions). This is approximately consistent with the differences found by RGH22 for the dust and gas temperature based modeling, and a similar result of a factor of $\sim 2$ difference was found at these $M_\mathrm{gas}$ values.

We also note a few additional differences.
We use similar grain-surface chemistry (as the reduced network was adopted from tests conducted in RGH22), but use the dust temperature to set our gas temperature whereas RGH22 computed the gas temperature.
We also do not include settling, while in RGH22 most of the $a\gtrsim 100\,\micron$ settles and plays a negligible role in the thermal balance, because the balance is dominated by the small grains that has higher density (Equation~\ref{eq:Tgas2}).
Another important difference is the dust composition used in our models vs. RGH22.  
In this work, we adopt a combination of astrosilicate and graphite based on WD01 -- that is similar to the dust composition used in \citet{Miotello2016, miotello_lupus_2017} -- while RGH22 used a mix of olivine (76\% by volume) and amorphous carbon (24\% by volume); more importantly, we construct the dust disk by fitting the SED of RU~Lup. 
How different dust compositions affect the disk structure, temperature and grain-surface chemistry, and how they could be better constrained are out of the scope of this paper and will be the subject of future work.

We find similar differences in the models for RU~Lup from \citet{miotello_lupus_2017}, although the dust composition used in our model is similar to theirs.
Their $M_{\mathrm{dust}}$ estimation was derived from the flux at mm-wavelength and not by fitting the SED, but the dust mass estimation of the two models converge to the same $M_{\mathrm{dust}} \sim 4\times10^{-4}\,M_\odot$.
However, their best-fit value of $M_{\mathrm{gas}}$ is a factor of $\sim 8$ smaller than the VHSE result and $\sim 14$ smaller than our Gaussian gas disk model estimate.
This can be partially attributed to the fact that the grid of Gaussian disk models used in \citet{miotello_lupus_2017} are not tailored to RU~Lup. 
For example, as noted earlier, the scale height parameters adopted impact the inferred line luminosity and therefore the mass estimate.
For the range of scale height parameters (together with other free parameters) used in \citet{miotello_lupus_2017}, the mass estimates in fact range from $4\times10^{-3}\,M_\odot$ to $4.8\times10^{-4}\,M_\odot$.
Another contributor is the $\mathrm{CO} \leftrightarrow \mathrm{CO_2}$ grain-surface chemistry conversion which is not accounted for in \citet{miotello_lupus_2017}; this could bring a discrepancy of a factor $\sim 2-3$ as noted by \citet{Trapman2021_CO2conversion} and RGH22.

We would like to note that there could be other processes at work that may deplete gas-phase CO at the surface, e.g., vertical diffusion of gas into the icy midplane where it may freeze out, although the extent to which this occurs will also depend on the ability of small grains to form and transport ices back into the surface layers \citep[e.g.,][]{Krijt_2020ApJ...899..134K, Powell_2022NatAs...6.1147P}. However, to correctly consider those processes require a full 2D transport model including the particle dynamics. Such simulations are not suitable for detailed modeling of observational data on individual targets, as they require knowledge of the disk's history; in fact, modeling presented here may help decipher disk conditions at different evolutionary stages from observations and inform the development of theoretical transport models.

In summary, the derived $M_{\mathrm{gas}}$ for RU~Lup in this work lies between the model grids from RGH22 and \citet{miotello_lupus_2017}, see Figure~\ref{fig6:LM-C18O-2-1}.
Our model is the first one that is specifically built for RU~Lup. We also fit the SED, \coo\ line spectrum, and radial distribution, while both previous models only matched the luminosity using a grid of models which resulted in larger uncertainties on the derived parameters. We thus demonstrate that \texttt{DiskMINT} is a promising tool for modeling individual disks and deriving more robust disk mass estimates.

\subsection{The Missing IR Emission in VHSE Models}
\label{subsec:discussIRSED}

\begin{figure}
    \gridline{\fig{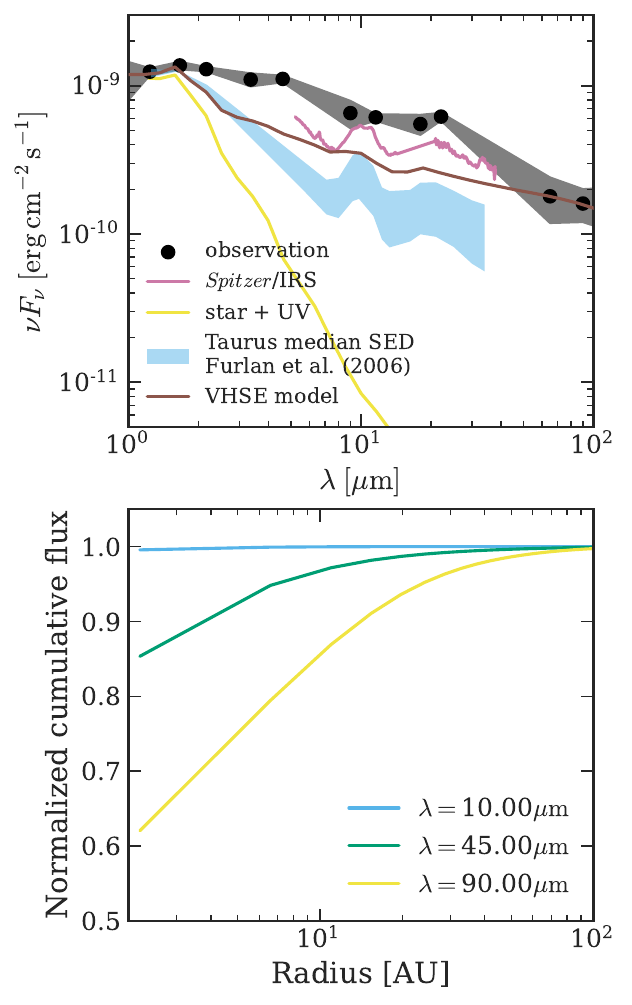}{0.45\textwidth}{}
            }
  \caption{
  Top panel: SED fitting results for the VHSE model. The same notations used in Figure~\ref{fig3:SEDandLineFitting} are adopted.
  The median SED from Taurus \citep{Furlan2006ApJS..165..568F} is added here with a blue shade. 
  Lower panel: normalized cumulative dust continuum fluxes for the best-fit VHSE models as a function of the distance from the star at different wavelengths. 
  Note that even at 90\,\micron{} 80\% of the flux arises within $10\,\mathrm{AU}$.
  }
  \label{fig7:IRSEDandcumuFlux}
\end{figure}

As discussed so far, the VHSE model successfully reproduces the \coo\ observations including the line velocity profile and radial distribution.
While the VHSE model presented in this work is capable of fitting the entire SED of the average $\sim 1-3$\,Myr-old disk \citep[e.g., the median Taurus SED from][]{Furlan2006ApJS..165..568F}, and can also match all available continuum photometry of RU~Lup beyond $\ga 100\,\micron$, it underestimates the infrared emission from the disk of RU~Lup by a factor of $\sim 2$ between $\sim 2 - 60\,\micron$ (Figure~\ref{fig7:IRSEDandcumuFlux} upper panel).

We first check and confirm that this IR continuum underestimation does not affect the gas mass determination from the $\mathrm{C^{18}O}$ (2-1) line.
\texttt{RADMC-3D} simulations show that the IR continuum emission comes from within a radial distance of $10\,\mathrm{AU}$ (see the cumulative dust emission in Figure~\ref{fig7:IRSEDandcumuFlux} lower panel), but the $\mathrm{C^{18}O}$ line emission mostly arises from the outer disk radius (Figure~\ref{fig3:SEDandLineFitting} lower right panel). 
There is therefore a deficit of dust emission from within $\la 10$ AU, indicating a possible missing physical process in our simple disk models. 

This lack of strong IR emission in VHSE models has also been noted previously, e.g., \citet{Woitke2016} for T-Tauri stars and \citet{Davies_HD142666Modeling} for Herbig Ae/Be stars.
Moreover, RU~Lup has one of the strongest IR excesses, a factor of $\sim 2$ higher than the upper boundary of the Taurus median SED (Figure~\ref{fig7:IRSEDandcumuFlux}), a region of similar age to Lupus  \citep{Comeron2008_handbook_hsf2.book..295C, Kenyon2008_handbook_hsf1.book..405K}.

One obvious shortcoming of our VHSE models is that we ignore gas thermal processes that are important, especially at the surface of the disk at small radii. 
Here other heating processes -- notable stellar high energy X-ray and UV photons -- will heat the gas to higher temperatures.
When densities are high, gas and dust are better coupled which leads to more small dust at higher elevations, increasing the IR excess.

Another intriguing possibility is that small dust grains ($\sim \micron$ size) are uplifted by a wind in the inner part of the disk (\citealt{Pascucci2022arXiv220310068P} for a recent review on disk winds).
This would lead to hotter dust at a higher scale height and thus increase the IR emission \citep{Bans_2012_WindsModelApJ...758..100B}. 
A parametric wind and disk model has previously been used to fit the strong IR excess from an Herbig disk \citep{Fernandes2018_addedwind}.
Interestingly, RU~Lup has a well-known inner wind detected via optical forbidden lines \citep[e.g.,][]{Fang2018_Winds_ApJ...868...28F, Banzatti2019_Winds_ApJ...870...76B, Whelan2021_Winds_ApJ...913...43W}.
It is quite likely that the wind (if dense, i.e., $n \gtrsim 10^{6-7}$ cm$^{-3}$) can loft small amounts of dust to greater heights and can explain the factor of $\sim 2$ deficit in the IR excess we find with the VHSE models. 
The hypothesis of a wind lifting small dust and increasing the IR excess warrants further exploration. 

\section{Summary and Outlook}
\label{sec:summary}

We developed a dust temperature-based self-consistent vertical hydrostatic pressure equilibrium disk model, \texttt{DiskMINT}, to compute gas disk masses.
\texttt{DiskMINT} is a \texttt{Python} code built on \texttt{RADMC-3D} and \texttt{LIME} for the continuum and gas line radiative transfer, respectively; and it includes a reduced chemical network suggested in RGH22 to determine the \coo\ distribution.
With \texttt{DiskMINT}, we introduce a target-based approach to estimate the disk mass in considerable details, where we fit the SED and also the \coo\ line emission.

We further test it on RU~Lup, whose disk was previously inferred to have just over a Jupiter-mass gas disk ($M_{\mathrm{gas}} \sim 1.5\times10^{-3}\,M_\odot$) and a gas-to-dust mass ratio $\Delta_\mathrm{gd}$ of only $\sim 4$.
We show that our model can match the long wavelength portion of the SED, the total \coo\ (2-1) and (3-2) line luminosity as well as the \coo\ (2-1) velocity and radial profiles with an order of magnitude higher mass ($M_\mathrm{gas} \sim 1.2\times10^{-2}\,M_\odot$) and gas-to-dust ratio ($\Delta_\mathrm{gd} \sim 30$).
We also test a Gaussian vertical density distribution that fits the SED better from IR- to millimeter-wavelengths and considers $\mathrm{CO} \leftrightarrow \mathrm{CO_2}$ conversion.
We find this Gaussian model that can match the line luminosity with even higher gas mass ($M_\mathrm{gas} \sim 2.1\times10^{-2}\,M_\odot$) and gas-to-dust ratio ($\Delta_\mathrm{gd} \sim 52$) but consider its large vertical height unrealistic.
We also find that the VHSE model underestimates the IR SED ($\lambda \sim 2 - 60\,\micron$) by a factor of $\sim 2$, which may indicate the need for considering more detailed gas thermal balance in the inner disk and/or an inner dusty wind from RU~Lup.

With our target-based approach, the RU~Lup's estimated disk mass is better-constrained, and it is larger than the Minimum Mass Solar Nebula \citep{Hayashi1981PThPS..70...35H}.
The larger mass is more in agreement with the young-age, high accretion rate, large disk size, and a lack of strong radial substructures in the disk of RU~Lup.
Our derived $\Delta_{\mathrm{gd}}$ is just a factor of a few lower than the ISM value of $\sim 100$.
This may indicate the disk has lost some of its gas within $\la 1\,\mathrm{Myr}$, or alternately, that CO is depleted by a factor of few. 
If the CO is depleted however by a factor of $\sim 10$ for RU~Lup as suggested by \citet{Zhang2020ApJ...891L..17Z} for disks in Lupus star forming region -- based on the data by \citealt{Ansdell2016} and attributed by them to the coupling of physical and chemical processes -- then the true $M_{\mathrm{gas}}$ would be as high as $\sim 0.1-0.2\,M_\odot$. 
Given that the stellar mass of RU~Lup is $\sim 0.2-1.2\,M_\odot$, such a massive disk would be gravitationally unstable, and we, therefore, consider large depletion factors unlikely.

In summary, a better understanding of disk physics and evolution could be achieved by modeling target-by-target and obtaining better-constrained disk masses for more disks of different ages.
The procedure of fitting the long-wavelength portion of the SED in combination with the $\mathrm{C^{18}O}$ line emission demonstrated in this work could be easily implemented on other targets with sufficient photometric data.
The \texttt{DiskMINT} code is also released  \citep{Deng_DiskMINT_2023_zenodo_8117966} and available in the public repository\footnote{https://github.com/DingshanDeng/DiskMINT}, so that the community can extend this approach to other disks.

\paragraph{Acknowledgments}
The authors thank C.P. Dullemond for helpful discussions and assistance on building our wrapper based on \texttt{RADMC-3D}, thank J. Barnes, A. Youdin and the anonymous referee for helpful suggestions and comments. DD, IP and UG acknowledge support from the NASA/XRP research grant 80NSSC20K0273 which made this work possible. Support for MR's research was provided by NASA’s Planetary Science Division Research Program, through ISFM work package `The Production of Astrobiologically Important Organics during Early
Planetary System Formation and Evolution' at NASA Ames Research Center.






\bibliography{DiskMassRULup}{}
\bibliographystyle{aasjournal}

\end{document}